\documentclass[journal=nalefd,manuscript=letter,layout=twocolumn,citeautoscript]{achemso}

\setkeys{acs}{etalmode=truncate}
\setkeys{acs}{maxauthors=10}
\setkeys{acs}{articletitle=true}
\usepackage{amssymb}

\usepackage{amssymb}

\usepackage{graphicx}% Needed to include png/jpg/pdf figure files
\usepackage{amsmath}
\usepackage{bm}
\usepackage{color, soul}
\usepackage[normalem]{ulem}

\usepackage[firstpage]{draftwatermark}%only first page
\usepackage[breaklinks]{hyperref}%hyperef package does internal and external linking (labels, urls)
\usepackage[all]{hypcap}%addition to hyperref, it ensures figure is correctly displayed (and not only the caption) when link is clicked

\title{Phase Transition Induced Carrier Mass Enhancement in 2D Ruddlesden-Popper Perovskites}

\author{Michal Baranowski}
\affiliation{Laboratoire National des Champs Magn\'etiques Intenses, UPR 3228, CNRS-UGA-UPS-INSA, Grenoble and Toulouse, France}
\alsoaffiliation{Department of Experimental Physics, Faculty of Fundamental Problems of Technology, Wroclaw University of Science and Technology, Wroclaw, Poland}
\author{Szymon J. Zelewski}
\affiliation{Laboratoire National des Champs Magn\'etiques Intenses, UPR 3228, CNRS-UGA-UPS-INSA, Grenoble and Toulouse, France}
\alsoaffiliation{Department of Experimental Physics, Faculty of Fundamental Problems of Technology, Wroclaw University of Science and Technology, Wroclaw, Poland}
\author{Mikael Kepenekian} 
\affiliation{Univ Rennes, ENSCR, INSA Rennes, CNRS, ISCR (Institut des Sciences Chimiques de Rennes) - UMR 6226, F-35000 Rennes, France}
\author{Boubacar Traor\'{e}}
\affiliation{Univ Rennes, ENSCR, INSA Rennes, CNRS, ISCR (Institut des Sciences Chimiques de Rennes) - UMR 6226, F-35000 Rennes, France}
\alsoaffiliation{Univ Rennes, INSA Rennes, CNRS, Institut FOTON - UMR 6082, F-35000 Rennes, France}
\author{Joanna M.\ Urban}
\affiliation{Laboratoire National des Champs Magn\'etiques Intenses, UPR 3228, CNRS-UGA-UPS-INSA, Grenoble and Toulouse, France}
\author{Alessandro\ Surrente}
\affiliation{Laboratoire National des Champs Magn\'etiques Intenses, UPR 3228, CNRS-UGA-UPS-INSA, Grenoble and Toulouse, France}
\author{Krzysztof Galkowski}
\affiliation{Laboratoire National des Champs Magn\'etiques Intenses, UPR 3228, CNRS-UGA-UPS-INSA, Grenoble and Toulouse, France}
\author{Duncan K.\ Maude}
\affiliation{Laboratoire National des Champs Magn\'etiques Intenses, UPR 3228, CNRS-UGA-UPS-INSA, Grenoble and Toulouse, France}
\author{Agnieszka Kuc}
\affiliation{Helmholtz-Zentrum Dresden-Rossendorf, Abteilung Ressourcen\"{o}kologie, Forschungsstelle Leipzig, Permoserstr. 15, 04318, Leipzig, Germany}
\author{Edward P. Booker}
\affiliation{Cavendish Laboratory, J.J. Thomson Avenue, Cambridge CB3 0HE, United Kingdom}
\author{Robert Kudrawiec}
\affiliation{Department of Experimental Physics, Faculty of Fundamental Problems of Technology, Wroclaw University of Science and Technology, Wroclaw, Poland}
\author{Samuel D. Stranks}
\affiliation{Cavendish Laboratory, J.J. Thomson Avenue, Cambridge CB3 0HE, United Kingdom}
\author{Paulina Plochocka}\email{paulina.plochocka@lncmi.cnrs.fr}
\affiliation{Laboratoire National des Champs Magn\'etiques Intenses, UPR 3228, CNRS-UGA-UPS-INSA, Grenoble and Toulouse, France}
\alsoaffiliation{Department of Experimental Physics, Faculty of Fundamental Problems of Technology, Wroclaw University of Science and Technology, Wroclaw, Poland}
\begin{document}

%\title{Come on baby shake my cations}

%\SetWatermarkText{FIRST DRAFT PLEASE DO NOT DISTRIBUTE}%Text for watermark
%\SetWatermarkLightness{0.8}%How light/dark
%\SetWatermarkScale{1.0}%Font scaling

\date{\today}

\begin{abstract}

\textit{There is a variety of possible ways to tune the optical properties of 2D perovskites, though the mutual dependence between different tuning parameters hinders our fundamental understanding of their properties. In this work we attempt to address this issue for (C$_n$H$_{2n+1}$NH$_3$)$_2$PbI$_4$ (with n=4,6,8,10,12) using optical spectroscopy in high magnetic fields up to 67\,T. Our experimental results, supported by DFT calculations, clearly demonstrate that the exciton reduced mass increases by around 30\% in the low temperature phase. This is reflected by a 2-3 fold decrease of the diamagnetic coefficient. Our studies show that the effective mass, which is an essential parameter for optoelectronic device operation, can be tuned by the variation of organic spacers and/or moderate cooling achievable using Peltier coolers. Moreover, we show that the complex absorption features visible in absorption/transmission spectra track each other in magnetic field providing strong evidence for the phonon related nature of the observed side bands.} 
\end{abstract}

\maketitle

%\section{Introduction}

The inherent sensitivity of lead-halide perovkites to ambient conditions  \cite{asghar2017device} is the Achilles heel which currently prevents the deployment of their superior properties \cite{stranks2013electron, vorpahl2015impact, hutter2017direct, huang2017understanding, yang2017recent,stranks2015metal, park2015perovskite} in real world applications\cite{park2016towards}. The last few years have witnessed rapid development of numerous perovskite derivatives \cite{giustino2016toward, dou2015atomically, mao2018hybrid, soe2017new, pedesseau2016advances, chen20182d, leblanc2017lead, spanopoulos2018unraveling}, in an attempt to overcome the environmental stability issue. For example, 2D Ruddlesden-Popper perovskites have already demonstrated power conversion efficiency greater than 10\%, with significantly improved stability\cite{tsai2016high, blancon2017extremely, yan2018recent, chen20182d}. Ruddlesden-Popper halide perovskites are natural type I quantum wells formed by thin layers of halide perovskite separated by organic spacers layers, which act as barriers \cite{even2014understanding, pedesseau2016advances, cao20152d}. They are described by general formula  A'$_2$A$_{m-1}$M$_m$X$_{m+1}$ where A' is a monovalent organic cation acting as a spacer, A is a small monovalent cation (MA, FA, Cs), and  M a divalent cation that can be Pb$^{2+}$, Sn$^{2+}$, Ge$^{2+}$, Cu$^{2+}$, Cd$^{2+}$, etc., X an anion (Cl$^-$, Br$^-$, I$^-$) and m=1,2,3... is the number of octahedra layers in the perovskite slab. The hydrophobic nature of the organic spacers significantly increases the stability of these compounds promising an excellent long term performance in photovoltaic and light-emitting applications \cite{chen20182d, yan2018recent, gong2018electron, tsai2016high, blancon2017extremely, smith2018white}.

A unique feature of 2D perovskites is that their properties can be tuned in far more ways than in the case of 3D perovskite semiconductors. The band gap can be tailored by varying chemical composition of the inorganic part \cite{Lanty2014, Du2017} or by changing the thickness of the octahedral slabs, which significantly impacts the optoelectronic properties \cite{blancon2017extremely, Blancon2018, Katan2019}. In addition, many of the 2D perovskite properties can be tuned by an appropriate choice of the building blocks with a plethora of different organic spacers to choose from \cite{chen20182d, yan2018recent}. Varying the organic spacer modifies the dielectric environment of the inorganic slab, affecting the exciton binding energy and the band gap \cite{Katan2019, pedesseau2016advances, Hong1992}. Moreover, the length and cross section of the organic ligands affect the distortion of the inorganic cages, modifying the band structure and the electron-phonon interaction\cite{Knutson2005,smith2018white, straus2016direct, Ni2017, Cortecchia2016, booker2017formation, gong2018electron}.

\begin{figure*}[t]
\centering
\includegraphics[width=1\linewidth]{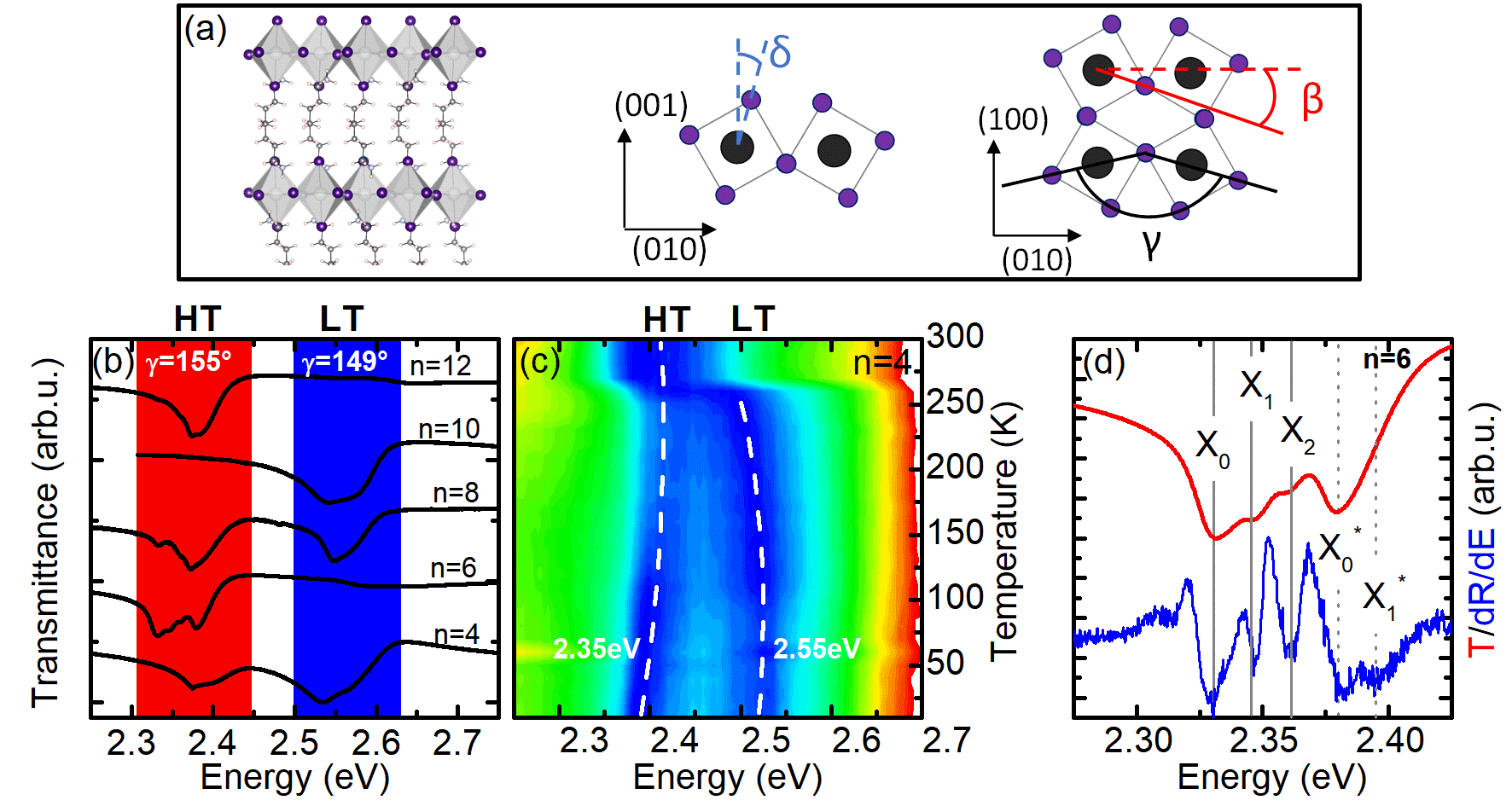}
\caption{(a) Scheme of (C$_4$H$_{9}$NH$_3$)$_2$PbI$_4$ crystal structure together with bridging Pb-I-Pb $\gamma$ angle and corrugation angle $\delta$ definitions. $\gamma=180-2\beta$ when $\delta=0$. (b) Transmittance spectra of (C$_n$H$_{2n+1}$NH$_3$)$_2$PbI$_4$ structures measured at $T=4$\,K. Spectra are labeled (4-12) according to number of carbon atoms in the organic spacer. Two groups can be distinguished, low energy transitions (HT phase) located at $\simeq2.35$\,eV and high energy transitions (LT phase) at $\simeq 2.55$\,eV which are characterized by $\gamma=155^\circ$ and $\gamma=149^\circ$, respectively. (c) Normalized transmission spectra of (C$_4$H$_{9}$NH$_3$)$_2$PbI$_4$ at different temperatures. Red corresponds to maxima and blue corresponds to minima in transmittance. The dashed white lines which follow the evolution of the two transitions are drawn as a guide to the eye. The LT phase appears around 270\,K, while the HT phase can be still observed down to $T=4$\,K. (d) Transmission and  derivative of reflectance spectra of (C$_6$H$_{13}$NH$_3$)$_2$PbI$_4$. Solid and dotted lines are separated by 15\,meV with respect to each other, indicating the position of the zero phonon lines and phonon replicas.}
\label{fig:transmition}
\end{figure*}

While the large variety of ways to tune the optical properties of 2D perovskites is undoubtedly their huge advantage, the interaction between different parameters hinders our fundamental understanding of their properties. To date their structural, dielectric, optical, and excitonic properties remain largely unexplored. Many of 2D perovskite family members exhibit complex absorption and emission spectra with many sidebands \cite{Kataoka1993, Xu1991, straus2016direct, Gauthron2010, Neutzner2018}. Today their origin is generally attributed to phonon replicas, \cite{straus2016direct, Ni2017, Neutzner2018, Gauthron2010} however, other interpretations invoking bound excitons or biexcitons can be found in literature \cite{Ishihara1990,fujisawa2004excitons, Hong1992}. Moreover, early magneto-optical studies report different diamagnetic coefficients for the various sidebands, which is not consistent with the phonon replica picture  \cite{Xu1991,Kataoka1993}. Another interesting aspect is related to the temperature driven phase transition \cite{Billing2007, Lemmerer2012, Billing2008, Ni2017}. The band structure of 2D perovskites is highly sensitive to octahedral distortion controlled by the organic spacer (templating)\cite{Knutson2005}. Therefore, a significant change of the effective mass and bandgap can be expected at the phase transition due to the rearrangement of the organic spacers in the different crystal phases \cite{Billing2007, Lemmerer2012, Billing2008}. Here, we address these issues for a series of (C$_n$H$_{2n+1}$NH$_3$)$_2$PbI$_4$ structures, schematically presented in figure\,\ref{fig:transmition} (a), with the varying length of organic spacers (n=4, 6, 8, 10, 12 is number of carbon atoms in the organic spacer). We use optical spectroscopy in magnetic fields up to 67\,T. The application of such high magnetic fields allows us to resolve reliably the diamagnetic shift, despite the large exciton binding energies in these systems (up to $\sim$ 400meV)\cite{Blancon2018, Yaffe2015}. We show that the complex absorption features precisely track one another in magnetic field, as expected for phonon replicas. The different diamagnetic shift of the transitions in different crystal phases is a direct proof of the significant modification of the carrier effective masses at the phase transition in agreement with detailed density functional theory (DFT) calculations. Since for some compounds the phase transition occurs close to room temperature, our findings suggest an additional way to engineer the properties of 2D perovskites via heating or a moderate cooling achievable using Peltier coolers.  

%\section{Results and Discussion}

\begin{figure*}[t]
\centering
\includegraphics[width=1\linewidth]{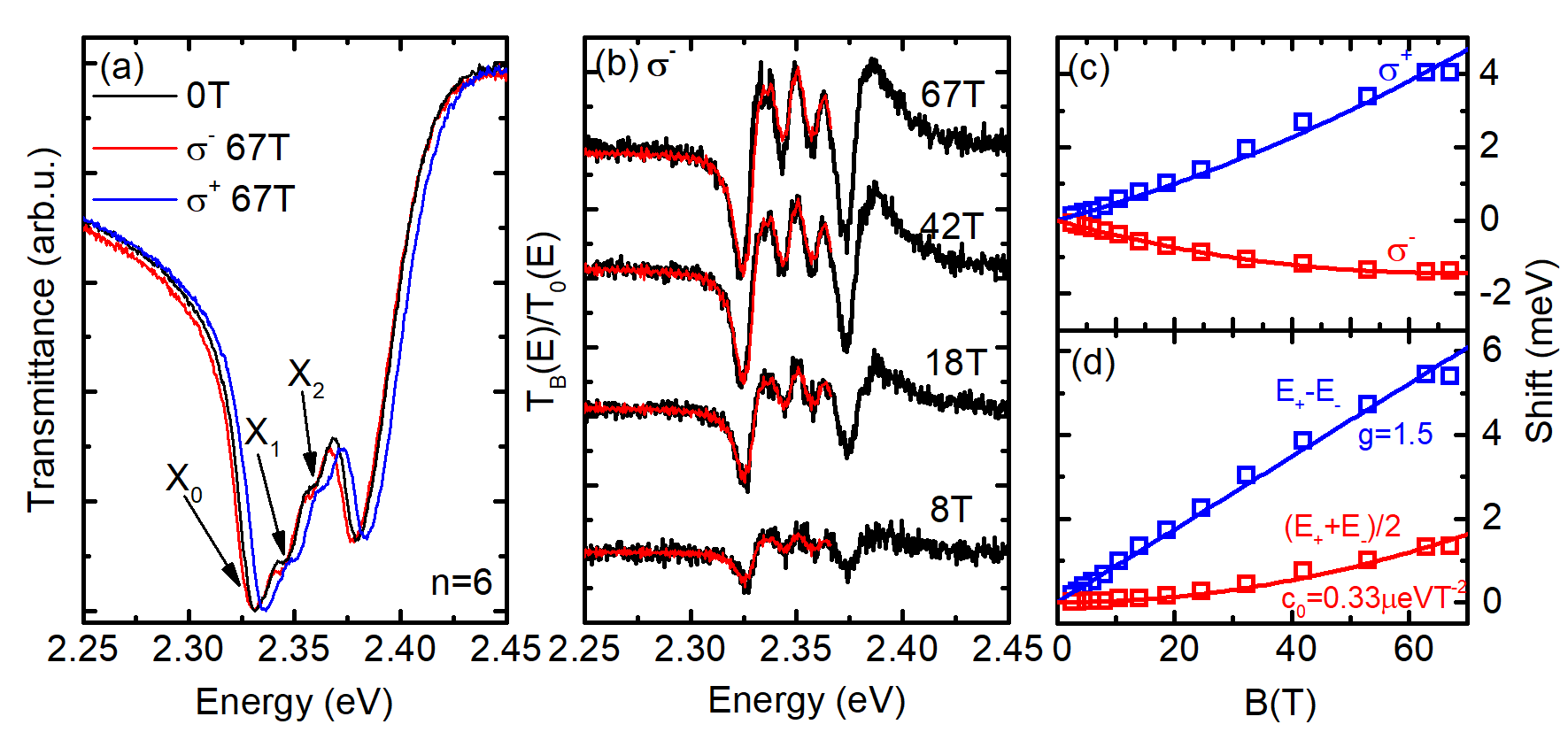}
\caption{(a) Comparison of transmittance spectra of (C$_6$H$_{13}$NH$_3$)$_2$PbI$_4$ sample at 0\,T (black) and 67\,T for $\sigma^+$ (blue) and $\sigma^-$ (red) polarization. (b) Transmittance spectra at different magnetic fields ratioed by 0\,T spectrum (black lines) and fitting results (red lines) for the energy range corresponding to X$_0$ transition and its phonon replicas. See detailed description in text. (c) Shift relative to 0\,T spectrum for $\sigma^+$ and $\sigma^-$ polarization and (d) extracted Zeeman and diamagnetic shift.}
\label{fig:fitting}
\end{figure*}

Figure\,\ref{fig:transmition} (b) shows a summary of the transmittance spectra of (C$_n$H$_{2n+1}$NH$_3$)$_2$PbI$_4$ with $n=4,\,6,\,8,\,10,\,12$ measured at $T=4$\,K. A strong excitonic absorption is observed in all cases. Transitions which occur around 2.55\, eV correspond to the low temperature (LT) 2D\,perovskite phase where the in plane Pb-I-Pb bridging bond angle $\gamma$ is $\simeq 149^\circ$ while transitions around 2.3\,eV correspond to the high temperature (HT) phase with a  $\gamma \simeq 155^\circ$ as revealed by our XRD analysis presented in SI and previous reports \cite{Billing2007, Lemmerer2012, Billing2008, Yaffe2015, Ni2017, sichert2019tuning} (see figure \ref{fig:transmition} (a) for angles definitions). The phase transition generally occurs in the vicinity of room temperature. However, the interaction between the spin coated 2D perovskites and the substrate can completely ($n=6, 12$) or partially ($n=4, 8$) inhibit the transition to the LT phase, the latter case leads to the formation of frozen high temperature phase domains\cite{Yaffe2015}. This effect can be directly observed in the temperature dependence of the transmittance spectrum of (C$_4$H$_{9}$NH$_3$)$_2$PbI$_4$ presented in figure\,\ref{fig:transmition} (c). The phase transition is observed around 270\,K as for single crystals\cite{Billing2007}. However, the signature of the HT phase remains even at $T=4$\,K due to the presence of frozen domains.     
   
The transitions in the transmittance spectra are asymmetric and in some cases multiple additional features (phonon replicas) are observed as shown in figure\,\ref{fig:transmition} (d) for (C$_6$H$_{13}$NH$_3$)$_2$PbI$_4$. In the transmission spectrum a series of dips is visible. The most pronounced one (X$_0$) is situated at $\simeq 2.33$\,eV and is followed by two phonon replicas (X$_1$ and X$_2$) at 2.345\,eV and 2.360\,eV separated by $\simeq 15$\,meV. A second transition at 2.38\,eV (X$_{0^*}$) is observed. These transitions are also observed in reflectivity. We plot the derivative of the reflectance spectrum where minima correspond to the transition energy\cite{yang2017unraveling}. The equal spacing of first three transition is characteristic for phonon-assisted transitions (phonon replicas)\cite{straus2016direct,Gauthron2010, pelant2012luminescence, Neutzner2018}. The energy of the involved LO phonon $\simeq 15$\,meV can be attributed to the vibration of the inorganic part of the 2D perovskite (Pb-I character) \cite{straus2016direct,Gauthron2010}. Interestingly, the energy of the phonon that dominates the absorption and emission spectra can be tripled by using a different organic spacer such as phenethylammonium\cite{straus2016direct,Gauthron2010,Neutzner2018}  showing the complex interaction between the inorganic and organic layers of 2D perovskites. The higher energy dip in transmission, labeled X$_0^*$ in figure\ref{fig:transmition} (c), is not at the correct energy, and has too large amplitude to be a phonon replica of X$_0$. It is most probably related to another phase of (C$_6$H$_{13}$NH$_3$)$_2$PbI$_4$ \cite{Ni2017} with a Pb-I-Pb bond angle, which remains close to $155^\circ$. For this transition, a weak phonon replica is observed in the reflectance derivative spectrum around 15\,meV above X$_{0^*}$.

\begin{figure*}[t]
\centering
\includegraphics{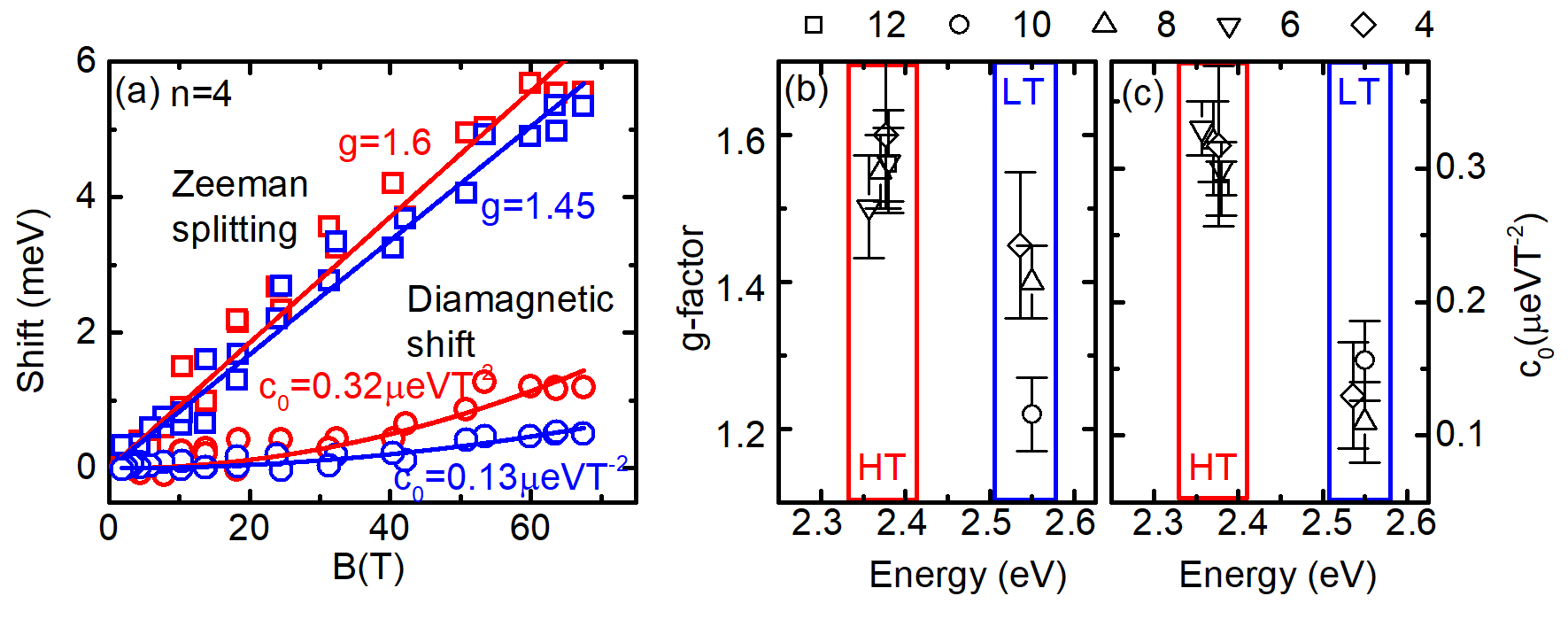}
\caption{(a) Comparison of Zeeman (open squares) and diamagnetic shifts (open circles) for the LT (blue) and HT (red) phases for (C$_4$H$_{9}$NH$_3$)$_2$PbI$_4$. (b) Measured g-factors and (c) diamagnetic coefficients as a function of transition energy. Symbols correspond to different number of carbon atoms in organic spacer.}
\label{fig:summary}
\end{figure*}

A critical test of the phonon replica picture is provided by their evolution in magnetic field. As the lattice vibrations are unaffected by the magnetic field, the phonon replicas should exactly track the evolution of the zero phonon line. Figure \ref{fig:fitting} (a) shows transmittance spectra at 0\,T and 67\,T  for two circular polarization $\sigma^+$ and $\sigma^-$. A small shift of a few meVs to higher and lower energies is clearly visible in the $\sigma^+$ and $\sigma^-$ spectra at 67\,T, respectively. 

%However, the complex shape of the spectra coupled with significant broadening complicates a quantitative determination of the %evolution with magnetic field. 

%Therefore we propose according approach for data analysis which allow for precise energy shift determination and prove that X$_0$, %X$_1$ and X$_2$ shift parallel in magnetic field. 

Figure\,\ref{fig:fitting} (b) shows the spectra measured in the magnetic field divided by the 0\,T spectrum. The ratioed spectrum has many sharp features (black curves). Their amplitude, width, and position depend sensitively on the energy shift with respect to the zero field spectrum.  Assuming the shape of the spectrum is unaffected by magnetic field, the ratioed spectra can be reproduced using a shifted 0\,T spectrum
\begin{equation}
\frac{T_B (E)}{T_0(E)}=\frac{AT_0(E+\Delta E)+C}{T_0(E)}
\label{equ:magic_fit}
\end{equation}
where $T_B (E)$ and $T_0(E)$ are transmittance spectra at a given magnetic field and zero field, respectively. $A$ and $C$ are constants which take into account any possible change of the transition amplitude or background during the field sweep. $\Delta E$ is the shift of the transition energy (with respect to zero field) at a given magnetic field. To determine the value of $\Delta E$ we fit the ratioed spectra from equation \eqref{equ:magic_fit} using the least-squares method (see red curves in figure\,\ref{fig:fitting}) (example of fitting procedure for other samples can be found in supplementary information). The agreement with experimental data is very good for all fields and results in the  shift error determination in the range of single percentage (see supplementary information). The fact that we can fit all features using the same energy shift unequivocally demonstrates that the transitions X$_0$, X$_1$ and X$_2$ exactly track each other in magnetic field, providing further support for the phonon replica picture. Importantly, this method allows us also to precisely determine the shift of the transition despite the complicated line shape. 

Figure\,\ref{fig:fitting} (c) shows the measured dependence of the exciton transition shift as a function of magnetic field for $\sigma^+$ and $\sigma^-$ polarizations.  In the Faraday configuration, the shift is determined by the spin dependent Zeeman energy and the exciton diamagnetic shift
\begin{equation}
    \Delta=\pm\frac{1}{2}g\mu_BB+c_0B^2
    \label{equ:shift}
\end{equation}
where the $g$ is the in-plane g-factor, $\mu_B$ is Bohr magneton and $c_0$ is the diamagnetic coefficient. The individual spin and diamagnetic contributions to the exciton shift are presented in figure \ref{fig:fitting}(d), where we plot the splitting and the average as a function of the field. Based on equation \ref{equ:shift}, $g \simeq 1.5$ and $c_0 \simeq0.33\,\mu$eVT$^{-2}$ for the lowest energy ($X_0$) transition of (C$_6$H$_{13}$NH$_3$)$_2$PbI$_4$.

\begin{figure*}[t]
\centering
\includegraphics[width=1\linewidth]{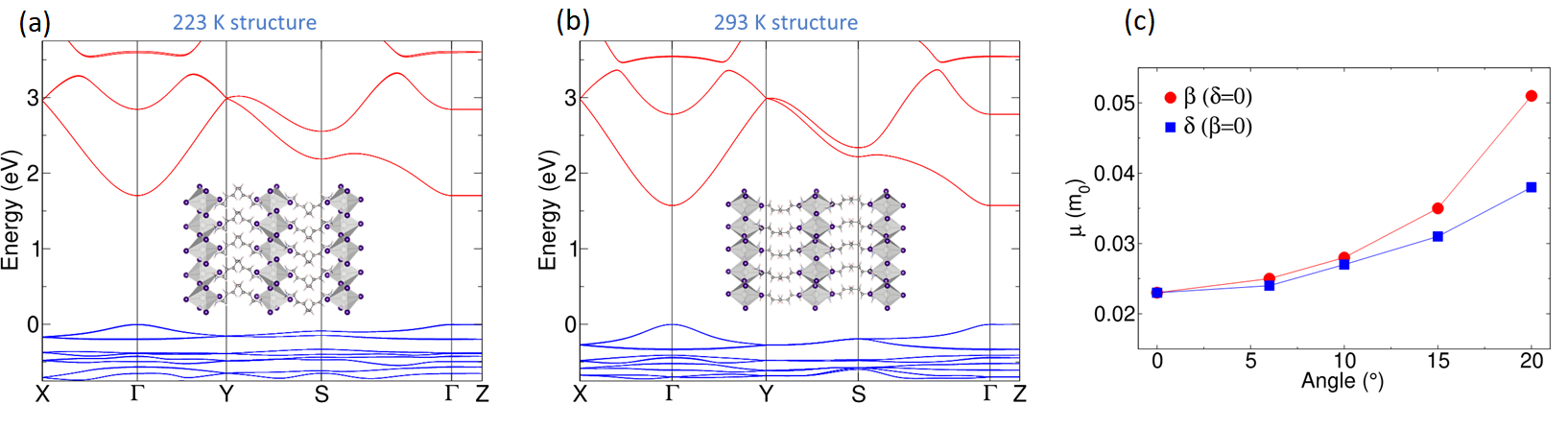}
\caption{Band structure corresponding to LT (a) and HT (b) phases of (C$_4$H$_{9}$NH$_3$)$_2$PbI$_4$ together with schematic crystal structure for high and low temperature phase). (c) Dependence of reduced exciton mass versus $\beta$ and $\delta$ angles.}
\label{fig:DFT}
\end{figure*}

The simultaneous observation of different crystal phases for $n=4$ and 8 allows us to investigate the electronic properties of both high and low temeprature phases of the 2D perovskite at low temperature. In semiconductors, an increase in the band gap (transition energy) is accompanied by an increase of the effective mass \cite{galkowski2016determination, galkowski2019excitonic, vurgaftman2001band}. This should change the diamagnetic coefficient which depends on the exciton reduced mass and wave function extension \cite{stier2018magnetooptics}. It is described by the general formula:
\begin{equation}
   c_0=\frac{e^2}{8\mu}\langle r^2\rangle 
\end{equation}
where $\mu$ is the reduced mass of the exciton and $\langle r^2\rangle$ is expectation value of the squared radial coordinate perpendicular to $B$. Since the extension of the exciton wave function decreases with increasing effective mass, the diamagnetic shift can serve as a sensitive tool for revealing the carrier mass variation. For example, in the simplest case of 2D and 3D hydrogen model the diamagnetic coefficient is proportional to $1/{\mu^3}$ \cite{miura2008physics}.  

An example of the Zeeman splittings and diamagnetic shifts observed in the $n=4$ sample are presented in figure\,\ref{fig:summary} (a). The Zeeman splitting for the LT ($g\simeq 1.45$ blue line) and HT ($g\simeq 1.6$ red) phases are comparable. The diamagnetic shift for the HT phase $c_0 \simeq0.32\,\mu$eVT$^{-2}$ is comparable to the HT phase in the $n=6$ sample. However, the diamagnetic shift of the LT phase $c_0 \simeq0.13\,\mu$eVT$^{-2}$ is about 2-3 times smaller. Similar trends can be observed for the range of investigated samples (see supplementary information for more examples) as shown in figure\,\ref{fig:summary} (b), (c) and \ref{tab:table}. The g-factors for all LT and HT phases are in the range of 1.2-1.6 in agreement with previous reports\cite{Xu1991,Kataoka1993}. The HT phase have diamagnetic coefficients grouped around $c_0 \simeq 0.3\,\mu$eVT$^{-2}$, while the LT phase has a diamagnetic coefficients which are systematically 2-3 times smaller. 
\begin{table}[]
    \centering
    \begin{tabular}{ |l|c|c|c|c|c| } 
        \hline
        n & 4 & 6 & 8 & 10 & 12 \\\hline
        LT $g$ & 1.45 & - & 1.4 & 1.22 & -\\\hline 
        HT $g$ & 1.6 & 1.5/1.56 & 1.55 & - & 1.56\\\hline 
        LT $c_0$ & 0.13 & - & 0.11 & 0.16 & -\\\hline 
        HT $c_0$ & 0.317 & 0.33/0.3 & 0.32 & - & 0.285\\\hline 
        \hline
 
    \end{tabular}
    \caption{Values of $g$ factors and diamagnetic coefficients $c_0$ ($\mu$eVT$^{-2}$) presented in figure \ref{fig:summary} (c).}
    \label{tab:table}
\end{table}

This observation strongly suggests a systematic mass enhancement in the LT phase. Within the 2D hydrogen atom model this requires a 30-35\% increase of the exciton reduced mass in the LT phase.

This rough estimation is supported by DFT band structure calculations performed over the LT and HT phases of (C$_4$H$_{9}$NH$_3$)$_2$PbI$_4$ (figure \ref{fig:DFT} (a) and (b)).
Both phases show a direct band gap at the center of the Brillouin zone with a band gap that slightly increases when going from the HT to the LT structure (1.57 eV to 1.70 eV) in good agreement with the experimental trend despite the well-known DFT underestimation of semiconductor band gaps.
As expected, these layered systems present flat bands in the stacking directions, whereas strong dispersions are displayed by the valence and conduction bands. Rough estimations of the hole and electron effective masses can then be extracted. More quantitative computation would require self-consistent many-body and relativistic treatments which are currently out of reach for such layered structures.~\cite{even2014d}
For the HT phase, we find hole ($m_h$) and electron ($m_e$) effective masses of $-0.103m_0$ and $0.049m_0$, respectively, leading to the reduced effective mass $\mu_{\text{HT}}=0.033m_0$. These values become $m_h=-0.200m_0$, $m_e=0.060m_0$ and $\mu_{\text{LT}}=0.046m_0$ for the LT phase. Therefore, we describe an increase of about 28\% of the reduced excitonic mass in the LT phase in good agreement with the experimental estimation.

The physical origin of mass change is related to the enhanced distortion of the octahedral cages, i.e., variation of bridging Pb--I--Pb bond angle and corrugation angle $\delta$, which is between the normal to the inorganic layers and the vector connecting the Pb atom and the terminal one (see the definition of angles in figure \ref{fig:transmition}).  The $\gamma$  ($\delta$) angles are about 155$^{\circ}$ (6$^{\circ}$) and 149$^{\circ}$ (12--13$^{\circ}$) for HT and LT phases respectively \cite{Billing2007, Lemmerer2012, Billing2008, pedesseau2016advances} regardless of the organic spacer length. The variation of this angle is crucial for the band structure. To visualize this we have performed calculations for an artificial 2D perovskite lattice structure as a function of distortion angles. For $\delta=0$ we define the angle $\beta$ as $\gamma=180-2\beta$. An increase of $\beta$ or $\delta$ results in the band gap opening accompanied by a flattening of the band dispersion (see also figures S10 and S11 in supporting information and previous studies \cite{Knutson2005,umebayashi2003electronic, pedesseau2016advances}). Therefore, the carrier mass carrier is enhanced as shown in figure\, \ref{fig:DFT} (c).  This explains also why the transition energy and the diamagnetic coefficients group into two sets for all investigate structures, in spite of the different length of organic spacers.

It is worth to note that in case of 3D MAPbI$_3$ a negligible change of the exciton reduced mass has been observed\cite{miyata2015direct} after the phase transition demonstrating the importance of the organic spacers for the band structure of 2D perovskites. With increasing octahedral slab thickness ($m>1$) of 2D perovskites the impact of the long spacers on the crystal structure is progressively decreased and the crystal structure is rather controlled by the MA cations. Therefore, it is expected that the band structures modification would not be as dramatic as in our case. However, this aspect still requires further experimental verification.

%\section{Conclusions}
In conclusion, using magneto-optical spectroscopy, we have shown that the complex absorption features in (C$_n$H$_{2n+1}$NH$_3$)$_2$PbI$_4$ are related to phonon replicas of excitonic transitions. The reduced mass of the exciton increases by around 30\% in the low temperature phase of (C$_n$H$_{2n+1}$NH$_3$)$_2$PbI$_4$. This is reflected by the 2-3 fold decrease of diamagnetic coefficient in the LT phase. The crystal structure can be controlled by the choice of appropriate organic spacers. Moreover our results suggest that temperature can be directly used as an additional parameter to engineer the properties of 2D perovskites. As the effective mass is a crucial parameter that determines carrier mobility in semiconductors, understanding how this quantity vary in complex 2D perovskite system is essential for identifying the impact on the properties that govern optoelectronic device operation. 

\section{Methods}

Alkylammonium iodide salts (RAI, RA = C$_n$H$_{(2n+4)}$N, n = 4, 6, 8, 10, 12) were prepared via neutralization of HI with RA. Unreacted species were removed by evaporation. The product was purified by recrystallization in minimal diethyl ether/excess hexane and isolated via vacuum filtration. Films were prepared by spin-coating solutions which were prepared by dissolving RAI and PbI$_2$ powders at a 2.5:1 molar ratio in a 1:0.34 volume ratio mixture of THF and methanol. Films were spin-coated from solutions of 20 mg/mL at 2000 rpm for 30 s and annealed for 15 min at 70 $^\circ$C. All precursors were purchased from Sigma Aldrich.
The crystals grow preferential with the lead iodide layers parallel to the glass substrate\cite{booker2017formation}

Magneto-transmission measurements were performed in pulsed magnetic fields up to 68\,T with a pulse duration of $\simeq 500$\,ms. The broad band white light was provided by a tungsten halogen lamp. The magnetic field measurements were performed in the Faraday configuration with the c-axis of the sample parallel to the magnetic field. The circular polarization was resolved in situ by a combination of a quarter waveplate and a polarizer, and the selection between $\sigma^+$ and $\sigma^-$ was achieved by reversing the direction of the magnetic field. The light was sent to the sample by optical fiber. The transmitted signal was collected by lens and coupled to another fiber. The signal was dispersed by a monochromator and detected using a liquid nitrogen cooled CCD camera. The sample was placed in a liquid helium cryostat. Temperature dependent transmission measurements were performed in the same setup. In such a configuration the probed area of the sample is in the range of few hundreds of $\mu$m$^2$. Nevertheless, the optical response related to the absorption is qualitatively and quantitatively the same as for thin films and single exfoliated flakes for 2D perovskite with thickness n=1 as was discussed in work\cite{blancon2017extremely} 
%PL measurements were performed in the back-scattering geometry in a closed cycle helium refrigerator. Samples were excited using a 405nm CW laser and the signal was detected using a liquid nitrogen cooled CCD camera. 

XRD was  performed  using a  Bruker  X-ray  D8 Advancediffractometer with Cu K$\alpha_{1,2}$radiation ($\lambda$= 1.541 \AA). Temperature control was provided by an Oxford Cryosystem PheniX stage. Spectra were collected with an angular range of 5$^\circ<2\Theta<60^\circ$ and $\Delta\Theta$= 0.01022$^\circ$ over 60min. Measurements were made on both as prepared spin coated films and powder samples obtained from the films. The Bruker Topas software was used to carry out analysis. More  detail can be found in ref. \cite{booker2017formation}.

First-principles calculations are based on density functional theory (DFT) as implemented in the {\sc SIESTA} package.~\cite{soler2002a, artacho2008a} Calculations have been carried out on experimental structures\cite{Billing2007} with the GGA functional in the revPBE form.~\cite{zhang1998a} Core electrons are described with Troullier-Martins pseudopotentials.~\cite{troullier1991a} The valence wavefunctions are developed over double-$\zeta$ polarized basis set of finite-range numerical pseudoatomic orbitals.~\cite{artacho1999a} In our calculations, spin-orbit coupling is taken into account through the on-site approximation as proposed by Fern{\'a}ndez-Seivane {\it et al.}~\cite{fernandez-seivane2006a} An energy cutoff of 150 Ry for real-space mesh size has been used and the Brillouin zone was sampled using a 5$\times$5$\times$1 Monckhorst-Pack grid.

\section{Supporting Information}
The Supporting Information is available free of charge on the ACS Publications website at DOI: XXXX
Detailed uncertainty analysis,  calculated band structures for varying $\beta$ and $\delta$ angles, summary of XRD structural characterization.     

\begin{acknowledgement}
This work was partially supported by the R{\'e}gion Midi-Pyr{\'e}n{\'e}es under contract MESR 13053031, BLAPHENE and STRABOT projects, which received funding from the IDEX Toulouse, Emergence program,  M.B. appreciates support from the Polish Ministry of Science and Higher Education  within  the  Mobilnosc Plus program (grant no.\ 1648/MOB/V/2017/0).
A.K. acknowledges ZIH Dresden for providing computational resources.
S.J.Z. also acknowledges the support within the Etiuda 5 scholarship from National Science Centre Poland (no. 2017/24/T/ST3/00257). 
\end{acknowledgement}

\begin{tocentry}
\includegraphics{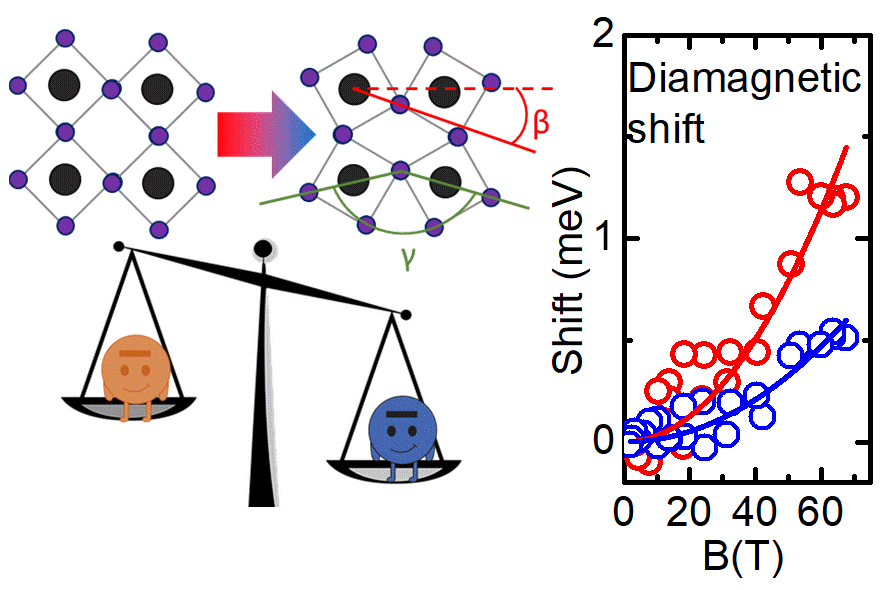}
\end{tocentry}

\bibliography{Bib}

\end{document}